\documentclass[opre,nonblindrev]{informs3}

\OneAndAHalfSpacedXI

\usepackage{setspace}
\usepackage[USenglish]{babel}
\usepackage[margin=1.16in]{geometry}
\RequirePackage{ifluatex}

\usepackage{fourier}
\usepackage{amsmath}
\usepackage{amssymb}
\usepackage{xspace}
\usepackage{amsfonts}%
\usepackage{changepage}
\usepackage{textcomp,marvosym}
\usepackage{cite}
\usepackage{multirow}
\usepackage{soul}
\usepackage{comment}
\usepackage{url}
\usepackage{booktabs}

\TheoremsNumberedThrough     
\ECRepeatTheorems

\EquationsNumberedThrough    

\MANUSCRIPTNO{MS-0001-1922.65}

\begin{document}



\RUNTITLE{Safer Workspaces at Times of Coronavirus}

\TITLE{Safer Workspaces at Times of Coronavirus: A Novel Use of Antibody Tests}


\ARTICLEAUTHORS{
\AUTHOR{Patricio Foncea}
\AFF{Operations Research Center, Massachusetts Institute of Technology, USA, 
	\EMAIL{foncea@mit.edu}}
\AUTHOR{Susana Mondschein}
\AFF{Department of Industrial Engineering, University of Chile and Instituto Sistemas Complejos de Ingenier\'ia, Chile 
	\EMAIL{susana.mondschein@uchile.cl}}
\AUTHOR{Ragheb Massouh}
\AFF{Medical advisor, GrupoBios, Chile 
	\EMAIL{rnmassou@uc.cl}}

}

\ABSTRACT{

The SARS-CoV-2 pandemic has transformed the way that the world functions.
 Health issues and population safety have driven countries' economies to a critical state; therefore, sustaining economic activity while keeping workers safe has become a worldwide goal.
In this paper, we present a novel safety protocol based on rapid antibody testing (ABT). Using discrete event simulation, we evaluated its performance on the cumulative number of infected workers, effective reproductive number ($R_e$), and active work force within a company.

Using a synthetic experiment, we showed that ABT twice a week (ABT 3) performed the best, detecting 5.7\% of infected workers, compared to 16.9\% when no ABT was applied. $R_e$ was reduced from 1.75 to 0.84, with a slight decrease in the active workers within the firm. A sensitivity analysis on the duration of the shedding period and sensitivity of ABT was performed and led to the same qualitative conclusions.
We applied this protocol in a Chilean winery: the estimation of the initial $R_e$ of 1.3 was reduced to 0.7 when the ABT 3 protocol was implemented, with a 27\% decrease in the number of infected workers.

Although ABT is not approved for COVID-19 diagnosis, our study shows that upgraded safety standards can already be implemented in workspaces.

}



\maketitle

%


\section{INTRODUCTION}

As SARS-Cov-2 spreads worldwide, governments 
struggle to keep people safe without collapsing the economy.   
Social distancing and quarantines have proven to be 
effective measures to save lives, yet their impact on the
economy has been experienced by countries worldwide.
The major challenge faced by many countries at this point of the pandemic is to find a way  to keep their critical industries, such as health, telecommunications, national security, transportation, food and energy, functioning and slowly reopen  nonessential industries,
while maintaining a safe environment for their workers.

In this paper, we propose a novel approach based on periodic SARS-CoV-2 antibody testing (ABT) to reduce the risk of contagion within the workspace. Using discrete event simulation,  we evaluate the impact of this approach using different protocols regarding testing frequency and the stage of the pandemic measured by the effective reproductive number, inside and outside the workspace. 
The use of ABT is the best available option at this point: it is inexpensive, fast, widely available, and easy to perform. We note, however, that the protocols analyzed in this study might be enhanced by using new and/or more effective  and inexpensive tests  as they become available. For example, the gold standard for diagnosis of SARS-CoV-2 is the RT--PCR. However, in most countries this is not a suitable option for screening: it is expensive and often scarce, the acquisition of the sample is very unpleasant, and delivery of the results is delayed by the required lab analysis. 
We also evaluated and discarded the use of point--of--care antigen tests due to their poor results to date.

The antibody rapid tests we consider require a simple finger prick blood sample to obtain a result within
15 minutes and have already gained  FDA-EUA (emergency use authorization) for their use. These tests can be taken at the workplace and have high specificity (higher than 95\%) and an 
acceptable sensitivity (approximately 90\%) after 7 days of symptom onset \cite{pan2020serological}, detecting the presence of specific anti-SARS-CoV-2 antibodies, which reflect concurrent or previous viral infection. 

We propose to apply ABT to the asymptomatic working population, assuming 
antibody responses similar to those described for symptomatic patients since some authors have seen 
that serum antibody levels do not necessarily correlate with clinical severity \cite{to2020temporal}. By applying this model, 
we intend to detect as early as possible an asymptomatic worker that could unknowingly be shedding 
the virus. Detecting and isolating the worker would prevent future spreading of the 
virus from this patient. Furthermore, this would 
allow the identification and quarantine of contacts, preventing further spreading of the virus.

Given ABT's sensitivity and specificity and the fact that antibodies develop gradually after infection, we propose  a conservative approach to quarantine potential infected workers. Thus, a positive test should be confirmed with an RT--PCR test. If this is not available, we recommend repeating the ABT test, and if it is positive again, consider the person as COVID-19 positive, despite  potential false positives. We also note that by using periodic testing, for example, twice a week, if a person with consecutive negative ABT results were to subsequently receive a positive antibody result, then they would have recently been infected, and therefore could potentially be infecting other coworkers. Although the specificity of ABT is not 100\%, the likelihood of a false positive largely decreases after the first negative screening.
Current studies show that most false positives have been attributed to an underlying cause, such as previous infection by other viruses from the Coronaviridae family; see, for example,  \cite{CDC}, \cite{Baraniukm2284}. This was also noted on previous cases of cross reactivity between corona viruses \cite{patrick2006outbreak}. The behavior of serologic tests in more specific conditions, such as autoimmune diseases or immunosuppressive therapy, have not yet been elucidated.

There has been an increasing concern worldwide  regarding sanitary measures when a worker has 
been identified as positive for the virus that could lead to closing a factory or organization. We strongly believe that a 
responsible risk management strategy applying safety measures, 
similar to the approach we propose and study in this paper, would justify that sanitation
authorities react less harshly when positive cases are detected.

The aim of this study is to design and evaluate protocols for screening using ABT together with work shifts when possible, to reduce the risk of COVID-19 infections within organizations. Thus, the goal is to detect potential asymptomatic COVID-19 workers to stop  the spread of the disease among coworkers and track all close contacts that might be infected and not yet showing symptoms, either because they are at the incubation period or asymptomatic. We note that as in any strategy for risk management, there are costs involved when isolating noninfectious workers as a preventive measure. We consider those in the evaluation of the proposed protocols.

We propose several realistic protocols and evaluate them using a discrete event simulation, where the main sources of uncertainty correspond to i)  initial infection prevalence in the working population, ii) days until onset of  symptoms, iii) duration of symptoms, iv) duration of infection after symptoms subside, and v) transition probability from susceptible to exposed, given the number of infected workers as a function of time. The latter is modeled to approximate the dynamics of the standard SEIR compartmental model.

The metrics used for the protocol evaluation are the firm's effective reproductive number, the cumulative number of  infected workers as a function of time, and the daily percentage of active workers as a function of the total work force. We note that as in any strategy for risk management, there are costs involved when isolating noninfectious workers as a preventive measure, which are considered in the evaluation of the proposed protocols.

Our results show that the implementation of ABT and shift protocols has a significant impact on reducing contagion within the organization. The magnitude of this reduction  depends on the effective reproductive number for the general population and the specific number within the organization. We have applied this methodology to a large Chilean winery, where the protocol of ABT  twice a week has been implemented, with  weekly day and night shifts of 12 hours plus additional weekly rests. Using the maximum likelihood estimation and the plant's historic data of infected workers since the beginning of the pandemic in Chile, we first estimated the plant's effective reproductive number before protocols were implemented. 
For the estimation of  asymptomatic infected workers, we used the results obtained at the first ABT.  With this information and data reported in the literature regarding the evolution of the disease, we projected the reduction in the number of infections  within the company compared to the baseline in which no ABT protocols were in place. After one month of implementation, the actual results are consistent with our projections. We also developed an ``alert system'' for the company in which the protocols can be activated or deactivated according to the stage of the pandemic. 

Finally, we emphasize that the protocols that we study and evaluate in this paper are complementary to other recommendations on avoiding the spread of the virus at work, such as  social distancing, personal protective elements, home offices, shifts, no casino lunches and sanitation, among others.

\section{MATHEMATICAL MODEL}

In this section, we describe  what is known about COVID-19 in terms of incubation, symptoms, and shedding periods. 
We also focus on the contagiousness of the virus and the production of antibodies that can be detected to isolate a potentially contagious patient. Then, we describe the relevant parameters and random variables  that are considered 
for the design and evaluation of  protocols using antibody tests and work shifts to reduce the risk of contagion within an organization. Finally, we describe the main features of the discrete event simulation model used to evaluate the performance of the proposed protocols, in terms of the following metrics: i) effective reproductive number, ii) percentage of infected workers actively working as a function of time, iii) cumulative infected workers as a function of time, and iv) daily percentage of active workers as a function of the total work force.

\subsection{Evolution of the disease: description and assumptions}
\label{assump}

Given the insufficient data for asymptomatic patients, the timeline of viral behavior for these patients
is extrapolated from what the literature describes as mild cases. This assumption is based on studies that show similar characteristics between both types of individuals \cite{zou2020sars}. 

In most cases, it is not known when the person was exposed to the virus and became infected. However, there is wide consensus that 
a patient becomes contagious, on average, two days before showing symptoms \cite{aylwardw}, \cite{he2020temporal}. We assume 
that the time from exposure to the virus until the symptoms develop  (incubation period), given that the infected person shows
symptoms, is distributed according to a lognormal distribution \cite{lauer2020incubation}. 

Current studies suggest that mild cases experience a duration of symptoms between 7 to 14 days \cite{lauer2020incubation}, and most likely, this period would be shorter for oligosymptomatic patients, i.e., patients with almost unnoticeable symptoms. Therefore, we assume a COVID-19-like symptom period of random duration in the range from 5 to 10 days for oligosymptomatic patients \cite{aylwardw}.

There is consensus  that the viral shedding period,  i.e., the period when the infected person is contagious,  starts approximately 2 days 
before the onset of symptoms \cite{aylwardw}, \cite{he2020temporal}. However, there is controversy on how long this period lasts after symptoms subside.
In our study, we assume that contagiousness of oligosymptomatics lasts randomly from 5 to 10 days after the COVID-19-like symptoms disappear \cite{aylwardw}.
This period seemed reasonable given the WHO recommendation at the time that we began this project (14 days of home isolation after symptoms resolve for mild cases \cite{world2020home}). However, this has changed in light of recent evidence, which although not of the highest quality, has shifted the latest guidelines from 14 days to only 11 days after the onset of symptoms \cite{TTT}.  However, due to the lack of studies that show an estimate for this duration on oligosymptomatic patients, in Section \ref{results}, we perform a sensitivity analysis regarding the expected value of this random variable.

The study in \cite{zou2020sars} finds that an infected person has a higher viral load within the first 5 days from symptom onset and hypothesizes that this situation could lead to a higher contagious rate. However, currently, there is no high-quality evidence that confirms this hypothesis. Therefore, we assume a homogeneous contagious rate over 
time \cite{xiao2020profile}, \cite{yongchen2020different}.

The WHO-China Joint Mission Report \cite{aylwardw} found that 81\% of the infected population presents mild symptoms, while other 
authors have found that this percentage is 61\% for mild cases \cite{liu2020viral}. An article from 
the Imperial College COVID-19 Response Team, \cite{ferguson2020report}, estimated that only two-thirds of the infected 
present recognizable symptoms. Additionally,  another study on pregnant women showed that 80\% might carry the virus 
without symptoms \cite{sutton2020universal}. In a long-term care facility in the United States, 
56\% of those who tested positive with PCR
were asymptomatic \cite{kimball2020asymptomatic}. Thus, combining current available information, we assume
that 50\% of the infected population will develop the disease in a oligosymptomatic fashion, i.e., few or minor symptoms; see \cite{gao2020systematic}  for a systematic review of asymptomatic infections.

We chose not to 
consider a higher incidence as found for specific population studies (e.g., pregnant or long-term care facility) 
or suggested by recent antibody testing on random populations, 
since this might lead to overestimation.  
Identifying these groups is key for pandemic control because asymptomatic
patients and patients with very mild COVID-19 symptoms may not seek health care 
nor receive diagnosis, which leads to underestimation of the burden of COVID-19. 
We note that the oligosymptomatic might shed the virus just as symptomatic patients \cite{zou2020sars}, \cite{chang2020protecting}.

Currently, the quarantine in place in Chile works as follows: if a person shows mild symptoms and is confirmed by PCR test, he/she is quarantined immediately for 11 days after symptom onset.
We note that in the protocols designed in this study, we used a quarantine of 14 days, since it was the health protocol during the first months of the pandemic. This more conservative quarantine would only have a slight impact on the number of active workers in the company but a gain in terms of not having potentially contagious workers in the workspace.

We also consider that an asymptomatic person without a previous history of COVID-19
with a  positive antibody test -- either for IgM or IgG --\footnote{IgM and IgG are immunoglobulins that form part of our immune response against different microorganism. They are two of the various known antibodies that humans produce. IgM is usually associated with the acute phase of an infection, while IgG can be seen during the  acute phase and posterior to it, lasting longer in our system. It  may prevent future reinfection or prepare for a better immune response in case of reinfection. Antibody rapid tests are specific for the detection of anti-SARS-CoV-2 antibodies.} is potentially an infectious agent, and therefore is also quarantined. If during this period, he/she presents symptoms,  the quarantine is extended for another 14 days since symptom onset. In both cases, the person at the end of the quarantine is considered recovered, and thus, not contagious.

Given the relatively low prevalence of confirmed 
SARS-CoV-2 up to July 2020 in Chile (less than 2.0\%) and that the first reported case was last March, we decided to use a conservative approach to interpret the antibody results for the asymptomatic-under screening population, as follows. 

\begin{table}[!ht]
	\begin{center}
		\begin{tabular}{cclc}
			\toprule
			\bf {IgM	} & \bf{IgG} & \bf {Interpretation} & \bf{Viral shedding assumption.}\\
			\midrule
			$-$	& $-$	& Still has not come in contact with virus & No$^{(*)}$\\
			&              &     or is in antibody window period.&	\\
			$+$  & 	$-$	& Acute phase of infection	&Yes\\
			$+$  & 	$+$	& Acute phase of infection	&Yes\\
			$-$   & 	$+$	& Acute phase of infection$^{(**)}$& 	Yes\\
			\bottomrule
			\label{assumptions}
		\end{tabular}
	\end{center}
	\vspace{-4mm}
	\begin{small}
		(*) We know that a negative antibody  test does not rule out 
		viral shedding, but the only way to confirm that would be by 
		PCR testing (or a good antigen test) and that is not currently available. Therefore, we interpret this result as the equivalent of having citizens going to work without screening.
		
(**)  We decided to interpret this result as active infection only after 2 or more consecutive antibody tests show negative results, to increase our sensitivity despite losing specificity, but ensuring a safer return to work. This is under the consideration of the following: $i )$ we are testing asymptomatic patients that have no previous history of COVID–19; $ ii)$ many will have previous antibody tests for comparison because we are testing twice a week; therefore, we would be witnessing real seroconversion; $iii)$ negative IgM could be a false negative (this ranges from 5-15\%); and $iv)$ if at any of the first 2 tests, the person yields IgM(-) and IgG (+), that would be considered a previous and not an acute infection. Thus, during the first week of testing, we would be making the cut between those with serology suggestive of previous infection and those without.

	\end{small}
\end{table}

As stated in the Introduction, the model's aim is to ensure a safe work environment;
therefore, we develop an algorithm for the interpretation of the antibody test
results that misses the least active viral shedding workers 
despite the possibility of quarantining a nonshedding worker 
until he has his PCR done.

Notice that the interpretation of the antibodies' results that is explained above  for the oligosymptomatics under a screening population differs from the interpretation done in cases of known COVID-19 patients who afterwards are tested with antibodies. We remark that the latter are not the subject of this study. Regarding those who came in contact with the virus and after some time are tested with antibodies and show IgM (-) and IgG (+), to date there is insufficient evidence to confirm their immunity; nonetheless, there are some studies that suggest that reinfection is unlikely \cite{an2020clinical}, \cite{bao2020reinfection}. Thus, in the near future, when more knowledge is acquired, the data gathered from screening will provide valuable information for public health policy.

\subsection{Model Description }

In this subsection, we first describe the notation used for the parameters and random variables of the model for the evolution of the disease.  Then, we describe the protocols that we considered for the application of antibody  tests and workers' shifts. Finally, we present the discrete event simulation model used for the evaluation of these protocols.

\subsubsection{Random variables}

\begin{itemize}
	
	
	
	\item $\Theta$ = onset period; time until symptoms show up, given that the person will be symptomatic.
	\item $V$=   viral shedding period, with expected value equal to  $ \mathbf {E}(V)$.
	\item $d_s=$ duration of symptoms. 
	\item $d_{ps} $= duration of post-symptom contagion period.
	\item $N_S(t)$ = number of workers that are susceptible on day $t$.
	\item $N_E(t)$ = number of workers that are exposed on day $t$.
	\item $N_I(t)$ = number of workers that are infected and contagious on day $t$.
	\item $N_R(t)$ = number of workers that have recovered from the virus on day $t$.
	\item $N$ = working population, which is constant over the short-term planning horizon considered. Therefore, $N = N_S(t) + N_E(t) + N_I(t) + N_R(t).$
	\item $W_S(t)$ = number of workers that are susceptible and working on day $t$.
	\item $W_E(t)$ = number of workers that are exposed and working on day $t$. 
	\item $W_I(t)$ = number of workers that are infected, contagious, and working on day $t$.
	\item $W_R(t)$ = number of recovered workers working on day $t$.
	\item $W(t)$ = total number of workers  working on day $t$.  $W(t) = W_S(t) + W_E(t) + W_I(t) + W_R(t)$. 
	\item $P(t)$ = population in the geographic area where the company is located.
	\item $P_S(t)$ =  susceptible population at time $t$ in the geographic area where the company is located.
	\item  $P_I(t)$ = infected population at time $t$  in the geographic area where the company is located.
\end{itemize}

\subsubsection{Parameters}

\begin{itemize}
	\item $R_0$ = Basic reproductive number (expected number of cases directly generated by one case in a population where all individuals are susceptible to infection).  This number is estimated at the beginning of the pandemic and is a function of  
	$R_0 = \tau c \mathbf {E}(V)$,  where $\tau$ is the transmissibility  (i.e., probability of infection given contact between a susceptible and infected individual), $c$ is the daily average rate of contact between susceptible and infected individuals, and $ \mathbf {E}(V) $ is the expected  duration of the shedding period \cite{heesterbeek1996concept}. Several estimates have been reported in the literature; see, for example,  \cite{viceconte2020covid}, \cite{zhang2020estimation}.
	
	\item $R_e(t)$  = effective reproductive number for the population. This number varies as a function of time, depending on the implemented measures, such as quarantines, social distancing, diagnostic capacity, and traceability,  among others.

	\item $R_e^w(\pi)$ = Effective reproductive number within the company. It corresponds to the expected number of new cases directly generated by one infected worker within the company. This number is calculated as $R_e^w (\pi)= \tau_e^w c_e^w \mathbf {E}(V(\pi))$, where $ \tau_e^w $ is the  transmissibility among workers, which is highly dependent on the sanitary measures implemented inside the company, 
	$c_e^w$ is the daily average rate of contact between susceptible and infected individuals inside the company, and $\mathbf {E}(V(\pi))$  is the effective duration of the disease inside the company, calculated as the number of days an infected worker stays actively working before being isolated. This number depends directly on the protocol $\pi$ implemented in the organization (the set of protocols  is described in Subsection \ref{prot}).

	\item $S = \Pr(\mbox{IgM}^+ \;\mbox{or}\; \mbox{IgG}^+ | \mbox{Infected})$ =  sensitivity of combined antibody test.

	
	\item $ E = \Pr(\mbox{IgM}^-\;\mbox{and}\; \mbox{IgG}^-|\mbox{Susceptible})$ =  specificity of the test.
	\item $ p_s$ = Probability of becoming symptomatic, given that the person acquires virus.
	\item $\Pr(S) $ = probability of being susceptible at the beginning of the planning horizon. 
	\item $\Pr(E)$ = probability of being exposed and not contagious at the beginning of the planning horizon. 
	\item $\Pr(I)$ = probability of being infected and contagious at the beginning of the planning horizon. 
	\item $\Pr(R) $ = probability of being recovered  at the beginning of the planning horizon. Note that $\Pr(S) + \Pr(E) +  \Pr(I) + \Pr(R) =1.$
	\item $S_h$ = number of working hours in a shift.
	\item $p(\pi,t)$ =  daily probability of a worker becoming infected  on day $t$, if protocol $\pi$ is implemented. It is a combination between the probability of becoming infected outside and inside the company. We assume that the worker sleeps for 8 hours; therefore, he is exposed inside and outside the working place for $S_h$ and $(16-S_h)$ hours, respectively.

	$$p({\pi},t) =  \frac{R_e^w(\pi)}{ \mathbf {E}(V(\pi)) } \frac {W_I(t)} {W(t)} (S_h/16) + \frac {R_e(t)}{ \mathbf {E}(V) } \frac{N_I(t)}{N(t)}((16-S_h)/16) $$
	
\end{itemize}

\subsubsection{Protocols} \label{prot}

In what follows, we describe the protocols defined for their evaluation. Some of them include the use of antibody tests or working shifts or a combination of both. 

\begin{enumerate}
	\item  \textbf{Base}: This is the baseline protocol, where only workers that show symptoms and are confirmed by PCR are quarantined. They return to work 14 days after symptoms subside.
	
	
	\item \textbf{ABT k}: This protocol includes antibody tests every $k$ days for all active workers, i.e., those who are going to work. If the test is positive for either $IgM$ or $IgG$, a PCR test is performed and the worker is quarantined until the results are known. If the result is negative, then that worker goes back to work; otherwise, he continues in quarantine for 14 days unless he shows symptoms at some point. In 
	the latter case, he stays isolated for  14 days after symptoms subside. We use a parameter that specifies the average number of days that it takes to have the results of the PCR test. At the peak of the COVID-19 pandemic in Chile, the processing time for  PCR tests could easily take up to a week.


	\item  \textbf{Shift 14}: This protocol considers that workers are divided into two shifts of 14 days each. If a worker is infected and shows symptoms, then the regular quarantine applies for him, either if this happens within or not within his shift.
	
	\item \textbf{Shift 14 + ABT k}: This protocol is the combination of shifts of  14 days and the use of antibody tests for  workers that are in an active shift.
\end{enumerate}

\subsubsection{Discrete Event Simulation}

\begin{figure}[h]
	\centering
	\includegraphics[page=5, trim={0 0.7cm 12cm 0}, clip, scale=0.65]{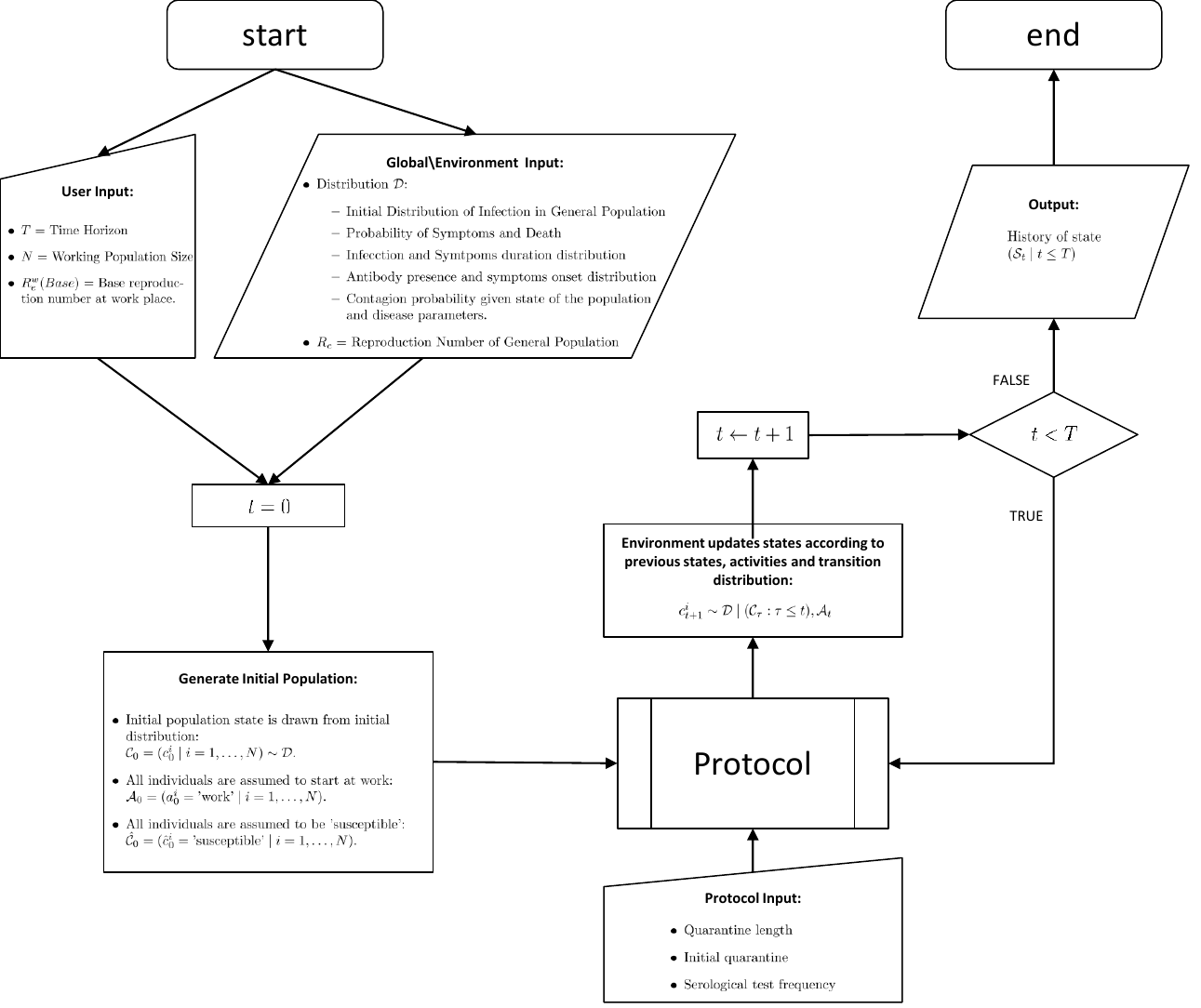}
	\caption{Flowchart of the algorithm.}
	\label{fig:flowchart}
\end{figure}

\begin{figure}[h]
	\centering
	\includegraphics[page=6, trim={0 3cm 18cm 0}, clip, scale=0.7]{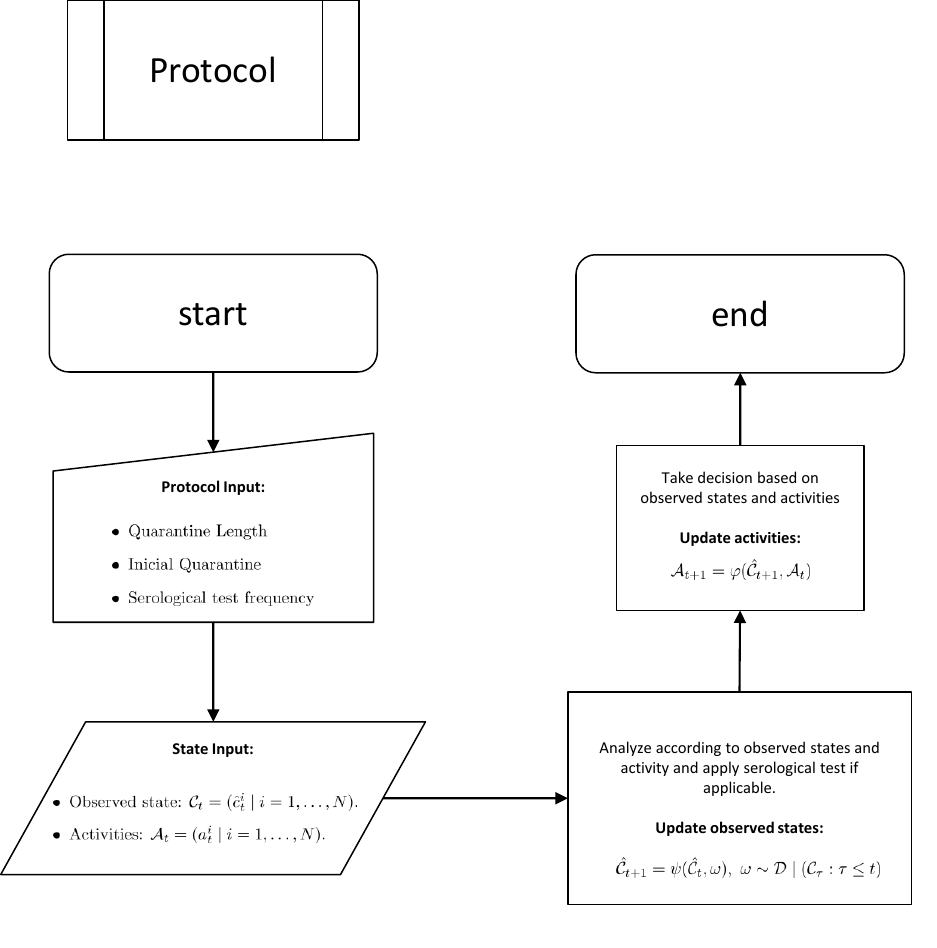}
	\caption{Flowchart of the Protocol.}
	\label{fig:flowchart_prot}
\end{figure}

The flowchart diagrams  in Figures \ref{fig:flowchart} and \ref{fig:flowchart_prot} describe the discrete event  simulation developed to evaluate the protocols designed to reduce the risk of infection within organizations.  Let $N=N_I(0) + N_S(0)+N_R(0)+N_E(0)$ be the total number of workers  at the beginning of the simulation and $T$ the planning horizon under study. To ease the notation, we assume that all random variables described in the previous section can be grouped in one joint distribution $\mathcal{D}$  -- although most of them are independent. The random variables considered in the simulation are as follows: i)  proportion of initial population at each compartment (S (susceptible), E (exposed), I (infected), R (recovered)), ii) days until symptom onset, ii) duration of symptoms, iv) duration of infection after symptoms subside, and v) transition probability from susceptible to exposed, given the number of infected workers.

For each individual (worker) $i=1,\ldots, N$, we denote by $c^t_i\in \{S,E,I,R\}$ his health state at time $t$, $\hat{c}^t_i \in \{S,E,I,R\}$ his observed state at time $t$ (for example, the decision maker might believe the worker is susceptible, although in reality, he is incubating the disease), and $a^t_i \in \{W,Q\}$ the working state at time $t$ (work or quarantine). For  protocol  \textbf{Shift 14}, we classify workers at home for 14 days; therefore, they can become infected as any other individual in the population. 

We encode a protocol by two functions $\psi$ and  $\varphi$. The first one, $\psi$ decides the observed health state of an individual. This  is based on any indication of illness the protocol is allowed to measure -- e.g., symptoms for Protocols 0, results of serological tests and symptoms for the rest. Note that given the stochastic nature of this indicator, the function $\psi$ takes as argument the vector of previous observed states, the current activity of each individual, and a random state drawn from $\mathcal{D}$ given the actual health state and activity, i.e., if he is quarantined or not. 
The second function, $\varphi$, decides whether the individual is allowed to work or must go into quarantine. It takes as input the vector of observed states and activities.

In each iteration, nature draws from $\mathcal{D}$ the new health state of the population given the history of health states up to that point and the activity of each worker. The distribution of workers in quarantine or work is relevant since it determines which workers can become infected by whom. Then, the protocol takes as input the current activities and the history of observed states, along with hyperparameters, such as duration of quarantine and frequency of serological testing if applicable, and uses the decision rules described above to update the activity of each individual.

After $T$ simulated days, the algorithm stops and outputs the full history of states and activities for every individual. Specifically, let $S_t^i=(c_t^i, \hat{c}_t^i, a_t^i)$ be the vector describing worker $i$; then, $\mathcal{S}_t=(S_t^i\mid i=1,\ldots N)$ and the algorithm outputs $(\mathcal{S}_t\mid 0\leq t\leq T)$.

\newpage

\section{Results} 
\label{results}

In this section, we describe the evolution of the disease for  the protocols described above and compare them in terms of the total number of infected workers, the average  daily percentage of active workers, and the effective reproductive number within the organization as a function of the implemented protocol. 
We also present a sensitivity analysis regarding two critical parameters: the duration of the contagion period and the sensitivity of the AB test. For the estimation of metrics, we used 5,000 simulations, with a coefficient of variation of less than 1.3\% for all cases.

\subsection{Data}

For illustration purposes, we consider an organization of 100 ``identical workers',' in terms of working hours,  close contacts, and sanitary conditions inside and outside the firm.  In the next section, we present a real application in which workers are tracked according to their own attributes.
In what follows, we describe the data used in the analysis presented in this section.

\begin{itemize}
	
	\item $R_e$ = 1.60
	
	\item $R_e^w(Base)$ = 1.75
	
	\item $S= 88\% $ (after a week since infection).
	
	\item $ E$ = 0.97
	
	\item $ p_s$ = 0.5
	
	\item $\Pr(S) $ = 0.983
	
	\item $\Pr(I)$  = 0.017
	
	\item $\Pr(R)$ = 0.00
	
	\item $S_h$ = 8  hours.
	
	\item $\Theta$ (days) = $\ln \mathcal{N}(\mu = 5.1,\sigma^2 = 1.8)$
	
	\item $d_s$ =  $U[7,10]$
	
	\item $d_{ps} = U[5,10]$
	
	\item $V$ =  $2+d_s+ d_{ps}$

\end{itemize}

\subsection{Comparison of Protocols}

\label{comparison}

In this subsection, we compare 6 protocols: \textbf{Base, ABT 7,  ABT 3, Shift 14, Shift 14 + ABT 7 }, and \textbf{Shift 14 + ABT 3}.  
Table \ref{tab: resumen} shows the performance of the baseline case and the 5 protocols evaluated for a horizon of 5 months and 100 workers; i.e., the number and percentage of  infected workers are equivalent.  As expected, the \textbf{Base} protocol performs the worst, having more than 8.5, 9.5, 11.2, 12.0, and 12.3 infected workers than having shifts of 14 days, antibody tests once a week or every 3 days, and the combination of 14-day shifts with antibody tests once a week or every 3 days, respectively. 
We note that the use of antibody tests significantly reduces the number of infected individuals, with a minimal impact on the average number of active workers: 98.5\% for \textbf{ABT 7} and 97.7\% for \textbf{ABT 3} compared to 99.0\% for the baseline case. We also note that the use of a combined strategy of antibody tests and shifts leads to the best results in terms of infected workers; however, it has a great impact on the number of active workers, reducing the latter to approximately a  half. When using 14-day shifts, antibodies every 3 days or once a week do not show an important difference; therefore, the less expensive protocol is recommended.

The last column of Table \ref{tab: resumen} shows the adjusted reproductive number within the company, $R_e^w(\pi)$, computed for each protocol. We recall that  $R_e^w (\pi)= \tau_e^w c_e^w \mathbf {E}(V(\pi))$, where we assume that the transmissibility $\tau_e^w$ and  the rate of contact $c_e^w$ among workers remain constant for all protocols; i.e., the sanitary measures do not change. However, the duration of the contagion period within the organization, $\mathbf {E}(V(\pi))$, strongly depends on the average number an infected worker stays working before being detected and isolated. Thus, in the baseline case, only symptomatic workers are isolated, leading to the highest $R_e^w(Base)$ of 1.95. We observe that the adjusted reproductive number greatly decreases when the protocols are in place, thus reducing the number of contagion days for a worker before being isolated. For example, in \textbf{Shift 14}, an asymptomatic  worker stays actively working for an average of a week before going home for his days off.

Figure 3 shows the evolution of the cumulative number of infected workers over time for a horizon of 5 months and 100 workers. For a longer planning horizon, we would eventually observe that the number of infected converge to a constant number. 

Finally, Figure \ref{fig:difference_ci} shows, for each simulation, the difference between the cumulative number of infected workers between the \textbf{Base}  and \textbf{ABT 3} protocols. The solid line corresponds to the average difference between the protocols,  and the area  under the upper curve envelope  contains 97.5\% of the simulation paths. We observe that by the fifth month, the difference of infected workers between the two protocols might reach up to 45 individuals, which in this case, corresponds to 45\% of the total number of workers in the company. Given the random nature of the process, the uncertainty and, therefore, the width of the band grows as a function of time.

We note that the characteristics of the pandemic evolve over time; therefore,  it also changes the effective reproductive number of the targeted population, $R_e(t)$. In section \ref{CCU}, where we discuss a real application of this protocol in a Chilean company, we develop a ``heat map'' with the benefits of the antibody test protocols as a function of the stage of the pandemic, $R_e(t)$.

\begin{table}[ht]
	\centering
	\begin{tabular}{ccccc}
		\hline
		
		Protocol  ($\pi$) & Total Infected Workers & \multicolumn{1}{p{2.5cm}}{\centering Daily Avg. Number of  Active Workers} & Adjusted $R_e^w(\pi)$ & \multicolumn{1}{p{2.5cm}}{\centering Avg. Viral Shedding Period $V(\pi)$ (Days)}\\ 
		\hline \hline
		\textbf{Base} & 16.9\% & 99.0\% & 1.75 & 10.5\\ 		
		\textbf{ Shift 14 }& 8.4\% & 50.1\% & 0.90 & 5.4\\ 
		\textbf{ABT 7} & 7.4\% & 98.5\% & 1.10 & 6.6\\ 
		\textbf{ ABT 3} & 5.7\% & 97.7\%  & 0.87 & 5.0\\ 
		\textbf{ ABT 7 + Shift 14} & 4.9\% & 49.8\% & 0.63 & 3.8\\ 
		\textbf{ ABT 3 + Shift 14} & 4.6\% & 49.5\% & 0.56 & 3.4 \\ 
		\hline
	\end{tabular}
	\caption{Comparison of performance among protocols for a horizon of 5 months.}
	\label{tab: resumen}
\end{table}

\begin{figure}[h!]
	\centering
	\includegraphics[trim={0 0cm 0cm 0}, clip, scale=0.55]{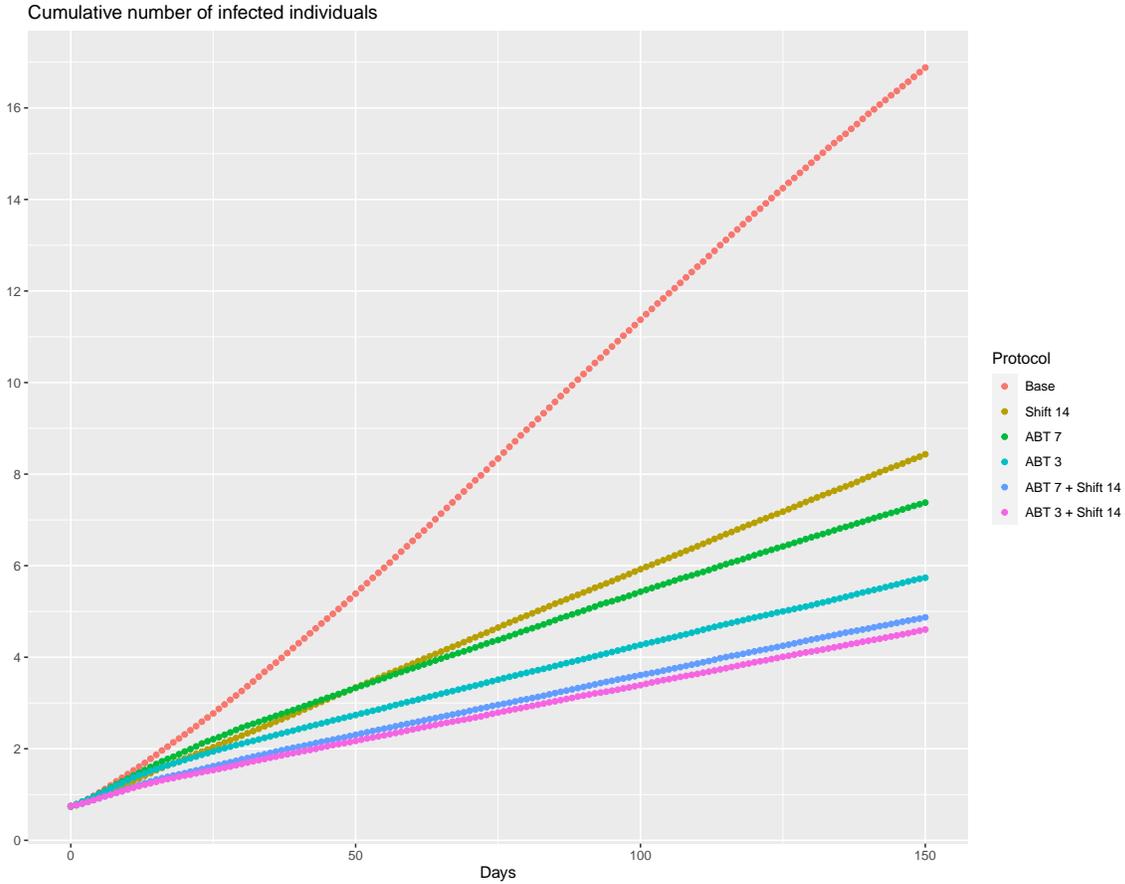}
	\caption{Evolution of cumulative number of infected workers for different protocols.} 
	\label{fig:cum_prot}
\end{figure}

\begin{figure}[h!]
	\centering
	\includegraphics[trim={0 0cm 0cm 0}, clip, scale=0.55]{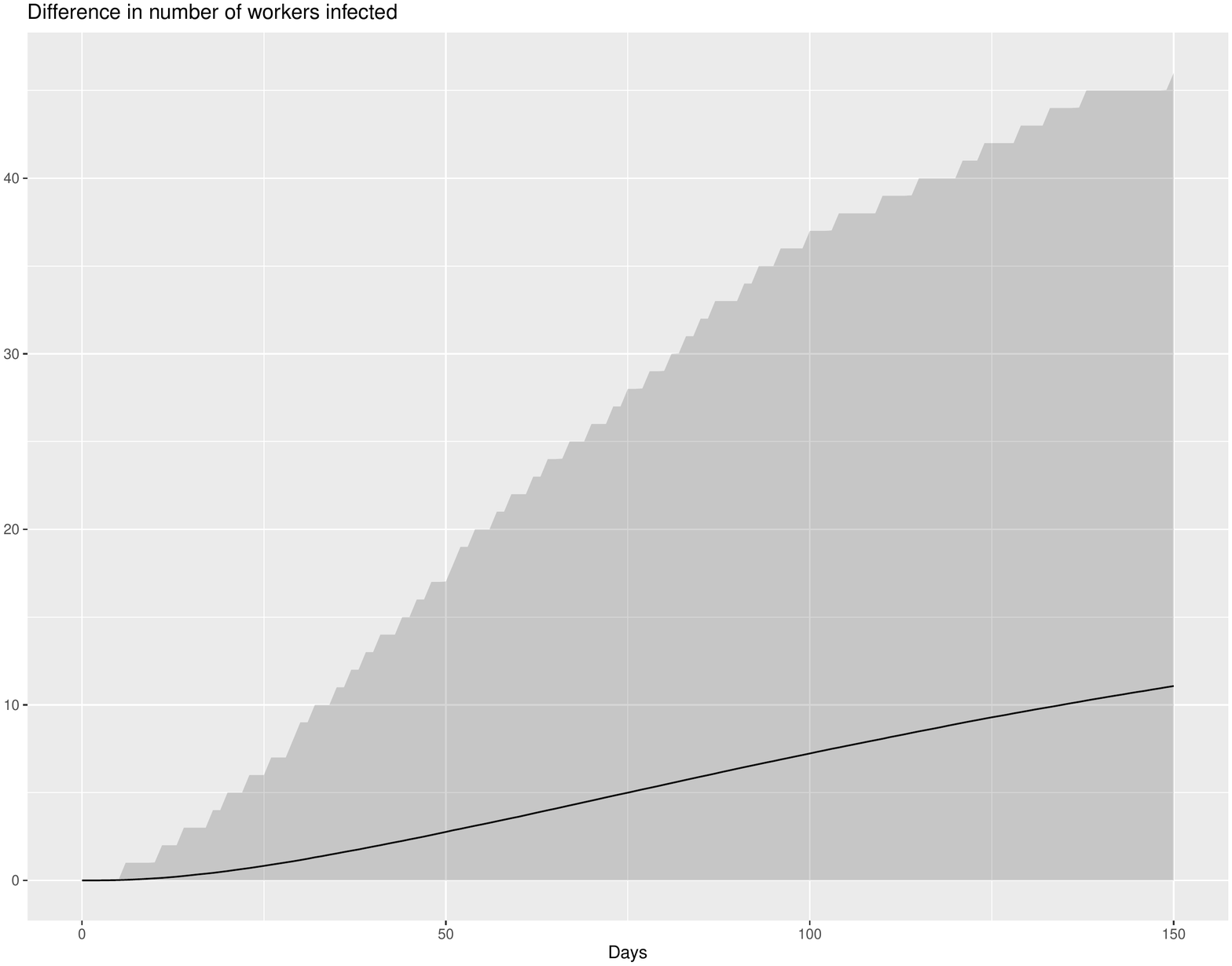}
	\caption{97.5\% of simulations are under the upper curve  for the difference of the number
		of infected workers between the \textbf{Base}  and \textbf{ABT 3} protocols. The solid line corresponds to the mean.}

	\label{fig:difference_ci}
\end{figure}

\subsection{Sensitivity Analysis}

In this subsection, we perform a sensitivity analysis regarding two parameters that might vary either because of different estimates reported in the literature or technical specifications: i) duration of the post-symptom contagion period,  and ii) sensitivity of the antibody tests. For this purpose, we simulate  the \textbf{ABT 3} protocol under different values for the sensitivity of the test and the duration of the post-symptom infection. 

\subsubsection{Sensitivity of antibody tests }

The sensitivity of the antibody tests varies depending on the manufacturer; therefore, it is important to understand its impact on the main metrics used to evaluate the protocols. In Table \ref{tab:sens_sensiv}, we report the total number of infected workers, the daily average of active workers, and the adjusted reproductive number within the plant in a horizon of 5 months. We observe that the sensitivity of the ABT has a negligible  impact on these metrics. The main reason for this interesting result lies on the periodicity of testing. Therefore, those infected and asymptomatic workers that could have been missed on one testing due to a false negative (one minus the sensitivity) have additional testing a few days apart (3 days in this example). Thus, with a sensitivity of 88\%, the probability of a false negative is 12\% on one test and decreases to 1.4\% after two consecutive tests. Similarly, with a sensitivity of 78\%, the probability of not identifying an infected worker after 2 consecutive tests decreases to 4.8\%. We remark that this is the main advantage of having a periodic ABT testing.

\begin{table}[ht]
	\begin{tabular}{c|cccc}
		\hline
		\multicolumn{1}{p{2.5cm}}{\centering Antibody Test Sensitivity} & Protocol ($\pi$) & Total Infected Workers & \multicolumn{1}{p{2.5cm}}{\centering Daily Average Number of  Active Workers}  & Adjusted $R_e^w(\pi)$ \\ \hline \hline
		N/A                 & Base     & 16.9\%                     & 99.1\%                            & 1.94   \\	
		0.98                & ABT 3  & 5.6\%                      & 97.8\%                            & 0.87           \\	
		\bf0.88                & \textbf{ABT 3} & \bf 5.7\%                      & \bf 97.7\%              &          \bf  0.87        \\	
		0.78                & ABT 3  & 5.7\%                      & 97.7\%                            & 0.90           \\
		0.58                & ABT 3  & 5.9\%                      & 97.7\%                            & 0.92           
		\\   \hline
	\end{tabular}
	\caption{Summary of the results for the \textbf{ABT 3} and \textbf{Base} protocols with different sensitivities of the AB test  \textit{after one week} of infection.  
	}
	\label{tab:sens_sensiv}
\end{table}

\subsubsection{Post-symptom contagion period}

Currently, there is a vast body of literature regarding the characteristics of the COVID-19 infection disease. However, as discussed in Subsection \ref{assump}, there is no agreement regarding the duration of the contagion period. Therefore, we perform a sensitivity analysis to study the impact of different durations for the contagion period on the metrics evaluated  above.  In the results reported  in Subsection \ref{comparison}, we consider a random post-symptom contagion period uniformly distributed between 5 and 10 days. In Table \ref{tab:sens_ips} we show how the metrics of the total number of infected workers,  daily average of active workers, and adjusted reproductive number within the plant change when shortening (lengthening) the post-symptom contagion period. 

We note that the basic reproductive number inside the company is constant and equal to 3, i.e., an infected worker infects, on average, 3 other workers if he remains in the workspace for the total duration of this disease. Thus,  the shorter the total shedding period is, the higher the daily infection rate. Therefore,  we observe that  for the Base and ABT 3 protocols, the  total number of infected workers and the adjusted reproductive number  decrease when the shedding period lengthens, with the corresponding increase in the daily number of active workers. 
 We also observed that in all cases for the duration of the shedding period,  the adjusted reproductive number is reduced to approximately a half when the \textbf{ABT 3} protocol is in place compared to that when there is no antibody test protocol.

We remark that in all cases,  the \textbf{ABT 3} protocol leads to significant decreases in the number  of infected workers compared to  the \textbf{Base} protocol, independently of the duration of the shedding period, in the analyzed range of values.

\begin{table}[h]
	\begin{tabular}{c|cccc}
		\hline
		\multicolumn{1}{p{2.5cm}} {\centering Post-symptom contagion period $(d_{ps})$}
		& Protocol ($\pi$)& Total Infected Workers & \multicolumn{1}{p{2.5cm}}{\centering Daily Average Number of  Active Workers} & Adjusted $R_e^w(\pi)$ \\ \hline \hline
		1 Day                          & ABT 3  & 10.5\%                      & 97.2\%                            & 1.14           \\ 
		& Base     								& 30.1\%                      & 98.2\%                            & 2.08         \\ \hline
		$U[1,5]$ Days                  & ABT 3  & 8.1\%                      & 97.5\%                            & 0.98          \\ 
		& Base    								& 22.6\%                     & 98.7\%                            & 1.90           \\ \hline
		\bf $\mathbf{U[5,10]}$ Days & \textbf{ABT 3}  & \bf 5.7\%                      & \bf 97.8\%                            & \bf 0.87           \\ 
		& \bf Base     							& \bf 16.9\%                     & \bf 99.1\%                            & \bf 1.75           \\ \hline
		$U[10,15]$ Days                & ABT 3  & 4.6\%                      & 97.9\%                            & 0.78          \\
		& Base     								& 16.2\%                     & 99.2\%                            & 1.68 \\  \hline    
	\end{tabular}
	\caption{Summary of the results for the \textbf{ABT 3} and \textbf{Base} protocols with different durations for the post-symptom infection period. 
	}
	\label{tab:sens_ips}
\end{table}

\section{Real case study: application of \textbf{ABT 3} to a Chilean winery}

\label{CCU}

In this section, we present an application of the antibody test protocol  \textbf{ABT 3} to  Vi\~na San Pedro Tarapac\'a S.A. (VSPT), Isla de Maipo plant. VSPT Wine Group is the second largest exporter of Chilean wine and is among the most important players in the Chilean market. The employees are classified into different categories such as administrative, professionals, and workers. Administrative, professionals, and some workers have shifts of 8 hours  from Monday to Friday, which we classify as administrative, while others rotate in three shifts: Monday to Friday day and night shifts and one week off (backups).  
Specifically, the composition of the work force is as follows:

\begin{table}[ht]
	\centering
	\begin{tabular}{lcr}
		\hline
		
		Shift & Schedule & Number of employees \\ 
		\hline \hline
		Administrative&  9 am  -- 5 pm & 202 \\ 		
		12 hours day shift & 9 am -- 9 pm & 51\\ 
		12 hours night shift & 9 pm -- 9 am & 47\\ 
		12 hours backup &   &  27\\ 
		12 hour shift &  9 am -- 9 pm & 18\\ 
		\hline
		Total &  & 345 \\ 
		\hline
	\end{tabular}
	\caption{Distribution of employees}
	\label{tab:employees}
\end{table}

The antibody test used, ActivaQ, has the following technical specifications reported by the manufacturer: i) a sensitivity and specificity of 99\% and 100\% for IgG antibody, respectively, and ii) a sensitivity and specificity of 88\% and 100\% for IgM antibody, respectively.
From April 6 to July 14, there were a total of 8 confirmed symptomatic cases. Additionally, after the first round of antibody tests at the beginning of the pilot, the presence of IgG antibodies was confirmed in a greater number of individuals, which correspond to workers who had the disease asymptomatically. Table \ref{tab:immune} shows the workers who have been infected and how their contagion was determined. It should be noted that all infections that were verified with antibody tests were classified as such during the first week of the application of the antibody test pilot.

\begin{table}[ht]
	\centering
	\begin{tabular}{lc}
		\hline
		
		Detection method & Number of employees at the beginning of pilot \\ 
		\hline \hline
		Symptomatic \& PCR(+)  & 8  \\ 		
		
		IgG(+) with close contact	  & 8 \\ 
		IgG(+) w/o close contact	 & 11 \\ 
		IgM(+)  \&  PCR(+) & 2 \\ 
		\hline
	\end{tabular}
	\caption{Distribution of infected workers at the beginning of the pilot.}
	\label{tab:immune}
\end{table}

If a worker has a positive IgM, a PCR test is taken, and  if this  is positive, the worker is quarantined.  Employees who have positive IgG and negative IgM are considered immune to the disease, for the time horizon of the pilot. However, since the period of immunity is not known, they are strongly recommended to follow all the sanitary measures of the company.
For the effective reproductive number of the population, we used the value reported in \cite{SaludPublica}. Symptomatic patients diagnosed in each town of the Metropolitan Region were used in this estimate. However, the incorporation of asymptomatic patients does not change this estimate, under the assumption that the proportion of symptomatic and asymptomatic patients does not change over time; see, for example, \cite{cori2013new}.

The  application  of the \textbf{ABT  3} protocol consisted of three steps: i) estimation of $R_e^w(Base)$ for the baseline case, ii) estimation of the average number of infected workers over time, since the beginning of the application of the protocol and the corresponding adjusted $R_e^w(Base)$ for this protocol, and iii) elaboration of a ``heat map''  to recommend the application of the protocol as a function of the stage of the pandemic and different values of the baseline reproduction rate $R_e^w(Base)$. 

\subsection{Estimation of the company's effective reproductive number for the baseline case}

Three months of historical data were available to perform the estimation of the reproductive number (or rate of infection) within the plant before the protocol was implemented. For each worker, we know which days she went to work (including absences due to in-house quarantine),  how many hours each day, and any \emph{COVID-19-related event}. These events include whether the worker has had a PCR test and its result, whether the worker presented COVID-19-like symptoms, or whether the worker was in close contact with someone who was infected. 

Following the same infection dynamics used in the simulation, we assume that each worker can become infected both inside and outside the plant. The probability of being infected outside the plant on a given day $t$, $q^n_t$, is computed as the product between the effective reproductive number at the town where worker $n$ lives \cite{SaludPublica},
and the COVID-19 incidence rate in that town, which is reported on a biweekly basis by the Ministry of Health \cite{MINSAL} . 

Inside the plant, the probability of being infected (which we denote $p_t(w)$) is proportional to the ratio between the number of infected workers currently at the plant over the total number of workers at the plant, which we call $i_t$. The proportionality constant corresponds to the effective reproductive number at the plant, $R_e^w(Base)$ -- which is assumed to be constant across time -- divided by the average number of days an infected worker stays infected in the plant, $V^w$, before being isolated. The former parameter is unknown and will be estimated, whereas the latter can be obtained through the data and equals $V^w=10.6$ days. 

The number of infected workers reported by the plant is likely to underestimate the real number, since the workers that were identified as COVID-19 positive are a subset of the ones that were tested, and the company only tested workers that presented COVID-19-like symptoms. That leaves out all possible asymptomatic or mild cases. To account for these cases, we use the results of the first two rounds of antibody tests performed during the protocol implementation to identify all workers that could have been infected with COVID-19 in the past.  If the worker had a COVID-19-related event, we assign the infection date randomly between 3 days before and 3 days after that event. For the case in which there is no COVID-19-related  event, we randomly assign it to any date during the observation period. Using the results of the first two rounds of  antibody tests, we detected 19 workers with a positive IgG and a negative IgM  antibody test. Therefore, we assume that these workers had the disease but they are no longer contagious and were not reported by the company in the list of infected workers. However, there is a chance that any of them could have been a false positive on repeated testing, for example, because of previous contact with other Coronaviridae virus, although this is very unlikely.

Finally, we use MLE to estimate the parameter $R_e^w(Base)$ with the infection history of the plant. Because we randomly assign infection dates for some workers, we perform multiple MLE estimations for 1000 realizations of the joint random variable that represent the infection date of these workers. We select the average of these estimators as our estimate of $R_e^w(Base)$. Specifically,  let $p_t^n$ be the probability that worker $n$ becomes infected on day $t$. We decompose this probability into two terms:
\[p_t^n = \alpha^t_np^w_t + \beta^t_nq_t^n.\]

Here, $p^w_t=R_e^w(Base)\cdot i_t/V_w$ and $q_t^n$, defined above,  are the probabilities that worker $n$ becomes infected during day $t$ inside and outside the plant, respectively, whereas $\alpha_n$ and $\beta_n$ are worker-specific nonnegative constants satisfying $\alpha^t_n+\beta^t_n\leq 1$, weighting the proportion of time the worker spends inside the plant and outside the plant, except for sleeping.

Then, the likelihood function to maximize is given by:
\[\prod_t\left(\prod_{n\in D_t} p_t^n\prod_{n\not\in D_t} (1-p_t^n)\right),\]
where $D_t$ is the set of workers that become infected on day $t$. Note that because of the random infection date assignments, $p_t^n$ and $D_t$ are random variables. The estimated $R_e^w(Base)$, computed as the average of the MLE for each random realization over 1000 scenarios, is equal to 1.3,  where an infected worker remains an average of 7.7 days. 

\subsection{Projections of the number of infected workers }

In this subsection, we estimate the number of  infections within the plant using the protocol of two antibody tests per week (Tuesday and Friday). These projections are compared with those resulting from continuing to operate with the current sanitary measures but without the inclusion of the \textbf{ABT 3} protocol. Finally, these projections are compared with the real results obtained during the application of the \textbf{ABT 3}  in the pilot plan carried out at the Isla de Maipo plant from July 14 to August 7.

The contagion projections were made through a discrete event simulation simulation, in which a daily monitoring of each  worker was carried out. A susceptible worker can be infected inside and outside the company, and this probability is calculated as a combination of both probabilities weighted by the fraction of the time he remains inside and outside  the company, on a daily basis. Therefore, this probability changes for those workers with administrative shifts, 12-hour day shifts, and with 12-hour shifts, day, night and back-up.
It should be noted that the probability of being infected outside the company is a function of the effective reproductive number in the town of residence (\cite{SaludPublica}). The probability of contagion within the plant is a function of the effective reproductive number in the plant, under the corresponding protocol.

\begin{figure}[ht!]
	\centering
	\includegraphics[trim={0 0cm 0cm 0}, clip, scale=0.55]{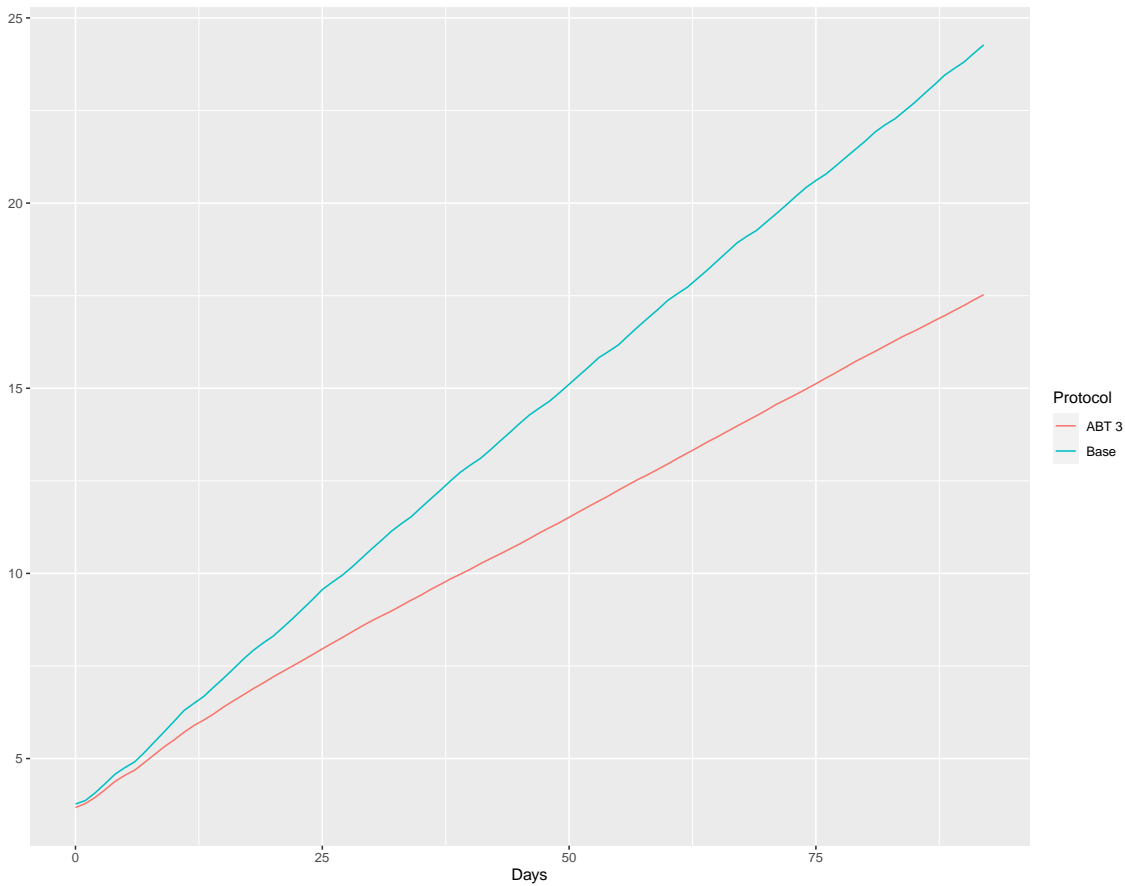}
	\caption{Projected number of infected workers under the Base protocol and ABT 3 protocol.} 
	\label{fig:abt_vs_base}
\end{figure}

Figure \ref{fig:abt_vs_base} shows  the projections of the total number of infected workers within the plant from July 14 to October 14 (3 months). It should be noted that at the beginning of the simulation we know that in practice, there were two asymptomatic infected workers who were identified with the antibody test and immediately isolated. However, in the baseline situation without an antibody protocol, these two patients would not have been identified and would remain in the plant, with subsequent potential contagion to other workers. The total number of infected as of October 14 in the base situation is 24.0 workers compared to 17.5 with the \textbf{ABT 3}, which corresponds to a reduction of 27\%.

Finally, we remark that the new effective reproductive number under the protocol equals 0.7, in contrast to the value of 1.3 that we had observed before the application of the antibody testing.

\subsection{Heat map for the ABT 3 protocol application}

In this subsection, we developed a heat map as a decision support system of when to suspend/activate the antibody test protocols.
These recommendations strongly depend on the stage of the pandemic; therefore, we developed a heat map that is a  function of the effective reproductive number in the general population as well as the one in the plant under the baseline case (i.e., with no antibody tests). We remark that the latter is a function of the sanitary measures in place as well as any additional changes in the worker shifts.

\begin{figure}[ht!]
	\centering
	\includegraphics[trim={0 0cm 0cm 0}, clip, scale=0.55]{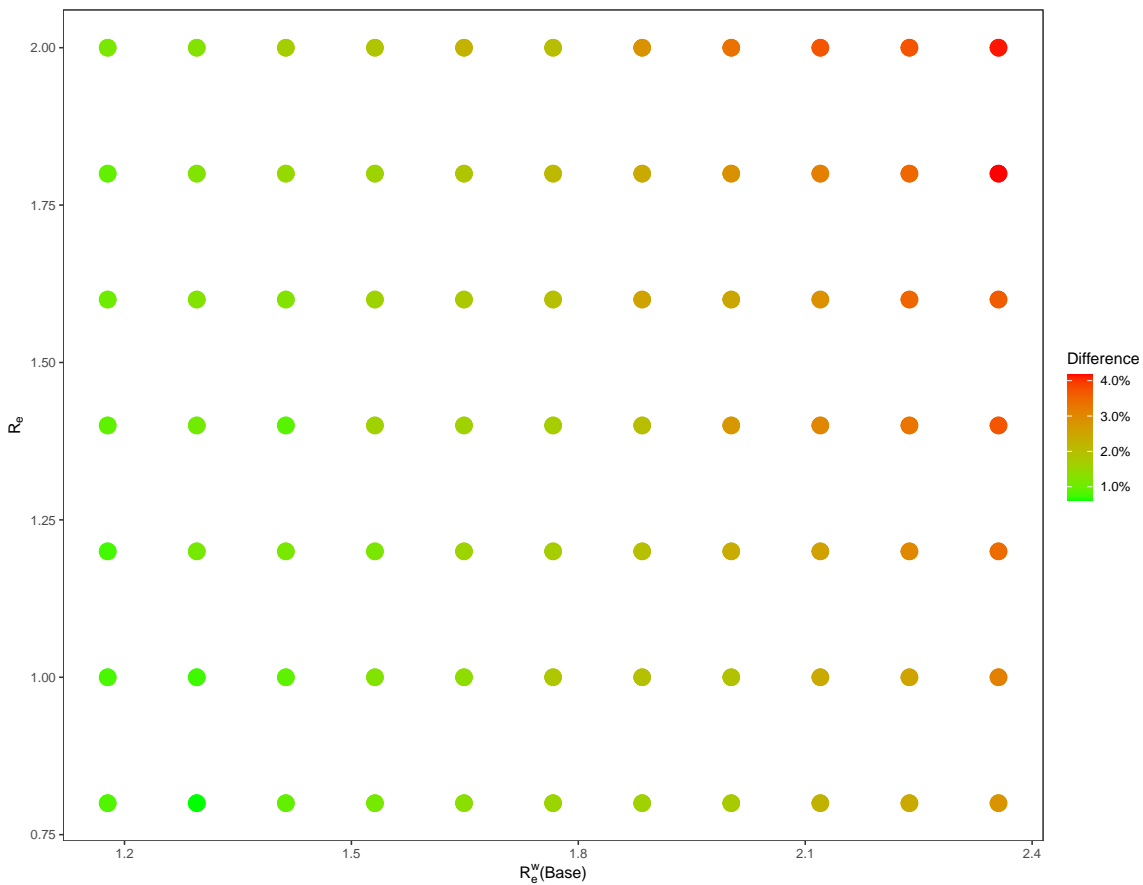}
	\caption{Heat map for using ABT 3 protocol as a function of the effective reproductive number inside and outside the firm. The color indicates the difference in infected workers -- as a percentage of the total workers -- between using ABT 3 and the Base protocols.} 
	\label{fig:heat_map}
\end{figure}

Figure \ref{fig:heat_map}
presents the combinations of $R_e $ and $R_e^w (Base)$ in green color for which the use of \textbf{ABT 3} leads to a decrease of less than 2\% in infections with respect to the \textbf{Base}  protocol. This percentage is calculated as the difference between infections with and without the \textbf{ABT 3 }protocol divided by the total number of workers in the company. Therefore, in the case of having 345 workers, this difference is less than 7 infected workers.
The graph shows the combinations of $R_e$ and $R_e^w (Base)$ in yellow for which the use of \textbf{ABT 3}  leads to a decrease greater than 2\% and less than 4\% in infections with respect to the \textbf{Base}   protocol. In the case of having 345 workers, this difference is greater than 7  and less than 14 infected workers.
The graph presents the combinations of $R_e$ and $R_e^w (Base)$ in red for which the use of \textbf{ABT 3}  leads to a decrease of more than 4\% in infections compared to the \textbf{Base}  protocol. In the case of having 345 workers, this difference is greater than 14 infected workers.

\section{Conclusions}

The application of the \textbf{ABT 3} protocol provides benefits in terms of avoiding contagion within the plant, since it is capable of detecting asymptomatic patients during the period in which they are still in the contagion phase and isolating them preemptively.
 Without the existence of these tests, asymptomatic patients would remain throughout their contagious period in the company.

Although antibody tests are only capable of detecting patients approximately a week after their contagion period begins, current evidence indicates that the contagion period can be between two or three weeks. Therefore, the effect of identifying and isolating these asymptomatic patients has a positive effect because otherwise they would continue, in some cases, to infect long after developing antibodies detectable by the test. With their isolation, the spread of the disease to other workers is prevented.

In our real case application, the effective reproductive number, defined as the number of infections that an infected worker produces within the company, is reduced from 1.3 to 0.7 with the application of the \textbf{ABT 3} protocol (2 times a week). The effect of the application of the \textbf{ABT 3} protocol has an important impact, reducing the value of the reproduction of the disease to less than one. This result implies that if the current conditions of the pandemic are maintained, the number of new infections within the plant should decrease over time.
For comparison purposes, we simulated  the application of the \textbf{ABT 1} protocol (once a week) and estimated that the effective reproductive number is 0.8 for this case. Therefore, it would be important to analyze the savings in monetary and nonmonetary costs when  taking only one antibody test per week, since the impact on the effective reproductive number is an increase of only 14\% with respect to that obtained with the \textbf{ABT 3}. 
The projection of the number of infections within the plant for the next 3 months, starting on July 14, shows that there is a reduction in the number of infections of 32\% and 28\% when \textbf{ABT 3} and \textbf{ABT1} protocols are applied, respectively (22 workers compared to 15 and 16). This reduction must be analyzed in the context of the health consequences that can derive from outbreaks or greater infections in the plant. Examples include the potential closure for 14 days when there are two or more weekly infections or quarantines of groups of workers due to close contact with an infected person. These sanitary measures have a significant impact on the productivity of the plant and should be considered when making the decision whether or not to continue with the \textbf{ABT} protocols.

It is important to emphasize that the estimates of contagion levels in the plant are dependent on the current situation of the pandemic in the country and, in particular, in the Metropolitan Region and are therefore valid as long as there are no significant outbreaks at the population level. If there are increases in infections in the population due to the suspension of quarantines, the number of infections within the plant will also increase because the probability of contagion outside the plant and its eventual spread within it increases.



\bibliographystyle{abbrv}
\bibliography{bibliography}

\end{document}